\documentclass[twocolumn]{webofc}
\usepackage[varg]{txfonts}   
\usepackage{graphicx}
\usepackage{color}
\usepackage{pifont}
\newcommand{\cmark}{\ding{52}}%
\newcommand{\xmark}{\ding{56}}
\newcommand{\msun}{M$_\odot$}
\newcommand{\lsun}{L$_\odot$}
\newcommand{\aap}{A\&A~}
\newcommand{\apj}{ApJ~}
\newcommand{\apjl}{ApJL~}
\newcommand{\mnras}{MNRAS~}
\newcommand{\aj}{AJ~}

\begin{document}
\title{NIKA2 observations around LBV stars}
\subtitle{Emission from stars and circumstellar material}

\author{\firstname{J. Ricardo} \lastname{Rizzo}\inst{1,3}\fnsep\thanks{\email{ricardo.rizzo@cab.inta-csic.es}} \and 
	\firstname{Alessia} \lastname{Ritacco}\inst{2} 
	\and
	\firstname{Cristobal} \lastname{Bordiu}\inst{1}
}

\institute{Centro de Astrobiolog\'{\i}a (CSIC-INTA), Ctra.~M-108, km.~4,
           E-28850 Torrej\'on de Ardoz, Madrid, Spain
\and
           Institut de Radioastronomie Milim\'etrique (IRAM), E-18012 Granada, Spain
\and
           ISDEFE, Beatriz de Bobadilla 3, E-28040 Madrid, Spain
          }

\abstract{
Luminous Blue Variable (LBV) stars are evolved massive objects, previous to core-collapse supernova. LBVs are characterized by photometric and spectroscopic variability, produced by strong and dense winds, mass-loss events and very intense UV radiation. LBVs strongly disturb their surroundings by heating and shocking, and produce important amounts of dust. The study of the circumstellar material is therefore crucial to understand how these massive stars evolve, and also to characterize their effects onto the interstellar medium. The versatility of NIKA2 is a key in providing simultaneous observations of both the stellar continuum and the extended, circumstellar contribution. The NIKA2 frequencies (150 and 260 GHz) are in the range where thermal dust and free-free emission compete, and hence NIKA2 has the capacity to provide key information about the spatial distribution of circumstellar ionized gas, warm dust and nearby dark clouds; non-thermal emission is also possible even at these high frequencies. We show the results of the first NIKA2 survey towards five LBVs. We detected emission from four stars, three of them immersed in tenuous circumstellar material. The spectral indices show a complex distribution and allowed us to separate and characterize different components. We also found nearby dark clouds, with spectral indices typical of thermal emission from dust. Spectral indices of the detected stars are negative and hard to be explained only by free-free processes. In one of the sources, G79.29+0.46, we also found a strong correlation of the 1mm and 2mm continuum emission with respect to nested molecular shells at 0.1 pc from the LBV. The spectral index in this region clearly separates four components: the LBV star, a bubble characterized by free-free emission, and a shell interacting with a nearby infrared dark cloud.
}
\maketitle
\section{Background}
\label{intro}
Luminous Blue Variable (LBV) stars are among the most massive objects, previous to core-collapse supernova (SN). The evolutive path around this short-lived stage (some $10^4$ yr) is not totally well settled yet, being considered previous to Wolf-Rayet \cite{gro14} and also direct precursors of SN \cite{gro13}. These stars develop strong and dense winds (up to $10^{-4}$~\msun), and are among the most intense sources, with luminosities up to several $10^5$~\lsun. LBVs are close to the Humphreys-Davidson limit \cite{gla93} and experience several dynamical instabilities which, at the end, result in violent outbursts of several solar masses. 

The environs of LBVs are then heavily affected by copious UV radiation and the mass-loss events. Almost all LBVs are surrounded by IR nebulae, composed by a rich mixture of dust and CNO-processed neutral and ionized gas \cite{oha03,miz10,bue17}. This circumstellar material (CSM) is therefore the consequence of ionization, shocks, dust formation and a presumably complex chemistry. The knowledge and characterization of the CSM linked to LBVs is therefore crucial not only to learn about the chemical and dynamical evolution of the Galaxy, but also to understand the physical processes which drive the massive star evolution.

Important efforts have been made in the study of the molecular structures around LBVs. The best studied object is probably G79.29+0.46, where we discovered a number of concentric, shocked shells associated with mass-loss events developed in the last $10^3-10^4$~yr \cite{riz08,jim10,riz14,pal14}. In {MGE042.0787+00.5084} (hereafter MGE042), we also found a molecular torus in slow expansion, with similar dynamical ages \cite{bor19a}. Recently, $\eta$~Carina was also the subject of molecular studies, where molecular material connected to the Homunculus and a more recent outburst were discovered \cite{loi12,loi16,mor17,smi18,bor19b}.

NIKA2 capabilities are exceptionally good to learn even more about LBVs environments at different scales \cite{ada18}. Its high sensitivity and the large field of view permit efficient and simultaneous observations of the stellar objects, their CSM and possible extended emission up to $\approx1$~pc from the star. These observations are made in two key bands, where both thermal dust and free-free emission of gas can compete. Under adequate conditions, NIKA2 even has potential to detect non-thermal (synchrotron) emission from this kind of sources.

In this paper we report the first results and preliminary analysis of an observational campaign in four fields around five confirmed or candidate LBVs.

\section{Observations}
\label{obs}
The observations were made at the IRAM 30m radio telescope, during the first open pool session of NIKA2 \cite{ada18} which has been carried out in October 2017, during three consecutive evenings. Four fields have been observed, enclosing 5 LBVs. All the observations were performed in day time, with typical precipitable water vapour of 2mm, opacity $\tau\approx 0.2$ measured at 225~GHz by the taumeter of the 30m telescope and stable weather conditions.

A summary of the observations is presented in Table \ref{tab-obs}. For each field, number of scans, integration time, typical elevations and {\it rms} are shown. Atmospheric opacity corrections are applied to the observed scans by using the NIKA2 skydip procedure described \cite{ada18}. This procedure uses NIKA2 itself as taumeter providing an opacity correction per observing scan. For all the scans, the opacities estimated varies between 0.17 and 0.24 for the 150~GHz band, and between 0.30 and 0.39 for the 260~GHz band.

\begin{table}[h]
\centering
\caption{Summary of observations}
\label{tab-obs}
\begin{tabular}{lccccc}
\hline
\multicolumn{1}{c}{Source} & scans & $t_{\mathrm{int}}$ & elv & $\sigma_{1\mathrm{mm}}$ & $\sigma_{2\mathrm{mm}}$ \\
 & & \footnotesize{hr} & $^\circ$ & \multicolumn{2}{c}{\footnotesize{mJy/beam}}\\
\hline
G79.29+0.46 & 12 & 1.1 & 68 & 5.2 & 1.3 \\
MGE042      & 60 & 2.0 & 56 & 1.6 & 0.5 \\
HD168625$^{(*)}$    & 24 & 1.0 & 36 & 2.1 & 0.6 \\
MGE027$^{(**)}$  & 30 & 0.9 & 48 & 2.7 & 0.8 \\
\hline
\vspace*{-5mm}
\end{tabular}
\begin{flushleft}
\footnotesize{
(*) HD168607 included in the same field of HD168625}\\
(**) MGE027 stands for MGE027.3839-003031 \cite{miz10}
\end{flushleft}

\end{table}

The maps were obtained by using NIKA2 IDL pipeline developed by the NIKA2 collaboration. In particular, we used an iterative method to subtract the correlated noise common to all the pixels of the arrays of NIKA2. At each iteration we combine all the maps for the different objects and we estimate the final averaged map. At the first iteration we subtract the noise estimated using all those pixels which are off-source, i.e. they detect the signal from the background which is a combination of atmospheric signal, electronic noise, cryogenic noise etc. Performing the difference between two consecutive maps we obtain the noise map that we subtract to the next iteration. At each iteration the signal to noise ratio increases, leading to a gaussian noise which is consistent with a white noise.

On-the-fly maps have been performed in different directions to overcome any striping effect that could appear along the scan direction. Combining the maps with different directions of scan we average this kind of effect and easy the offline data analysis processing.

\begin{figure*}[hbt]
\centering
\includegraphics[width=0.92\textwidth]{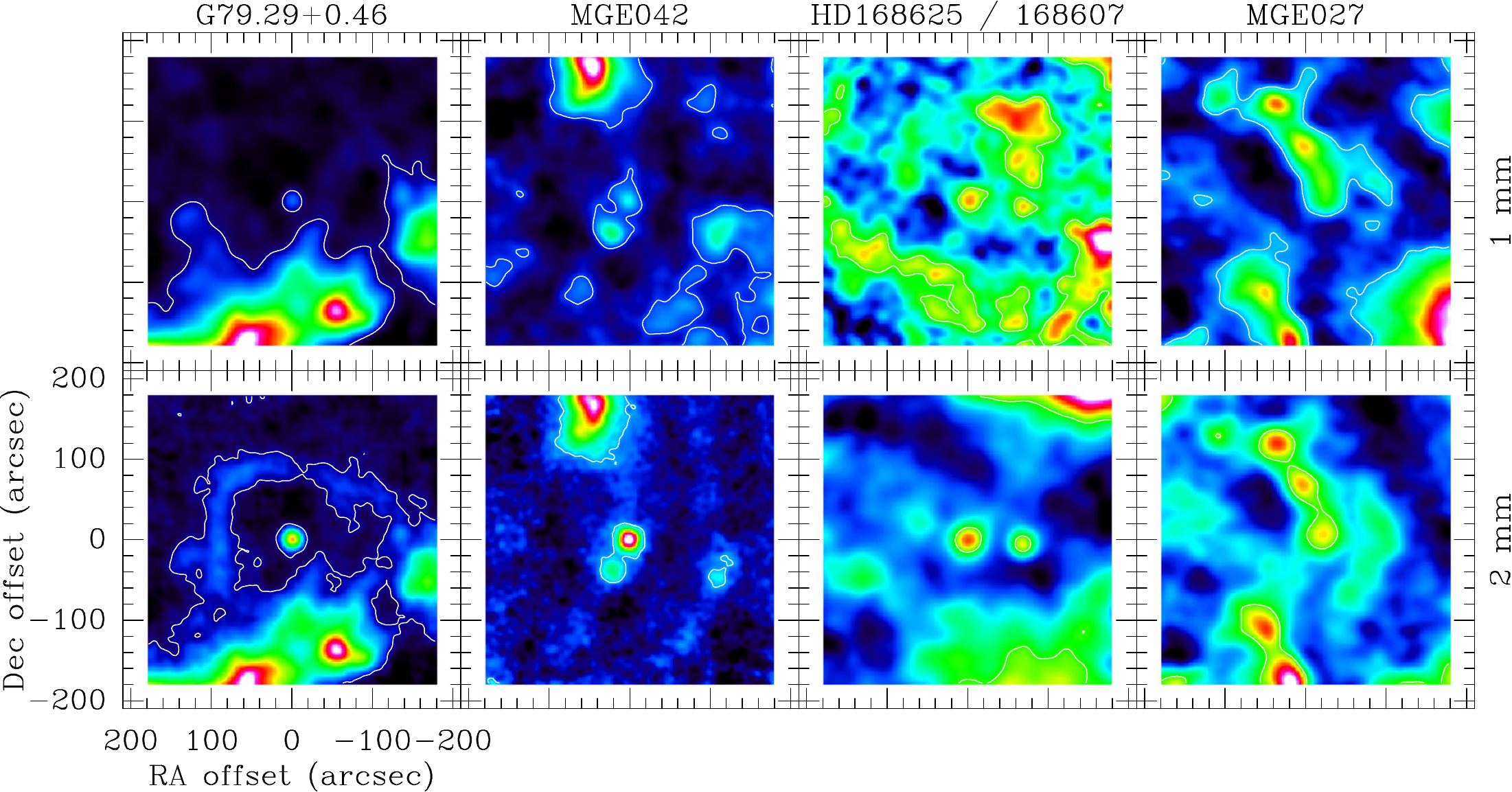} 
\label{bany}
\caption{Final images in the four fields observed, smoothed to a HPBW of $20"$. Top row contains the images at 260~GHz, while the bottom row depict those images at 150~GHz. Sources are indicated in the upper parts of the 260~GHz images. A diversity of features are seen, including the LBV stars themselves, but also some CSM and even Galactic dark clouds. White contour is at approximately 5-sigma level.}
\end{figure*}

\begin{figure*}[hbt]
\centering
\includegraphics[width=0.92\textwidth]{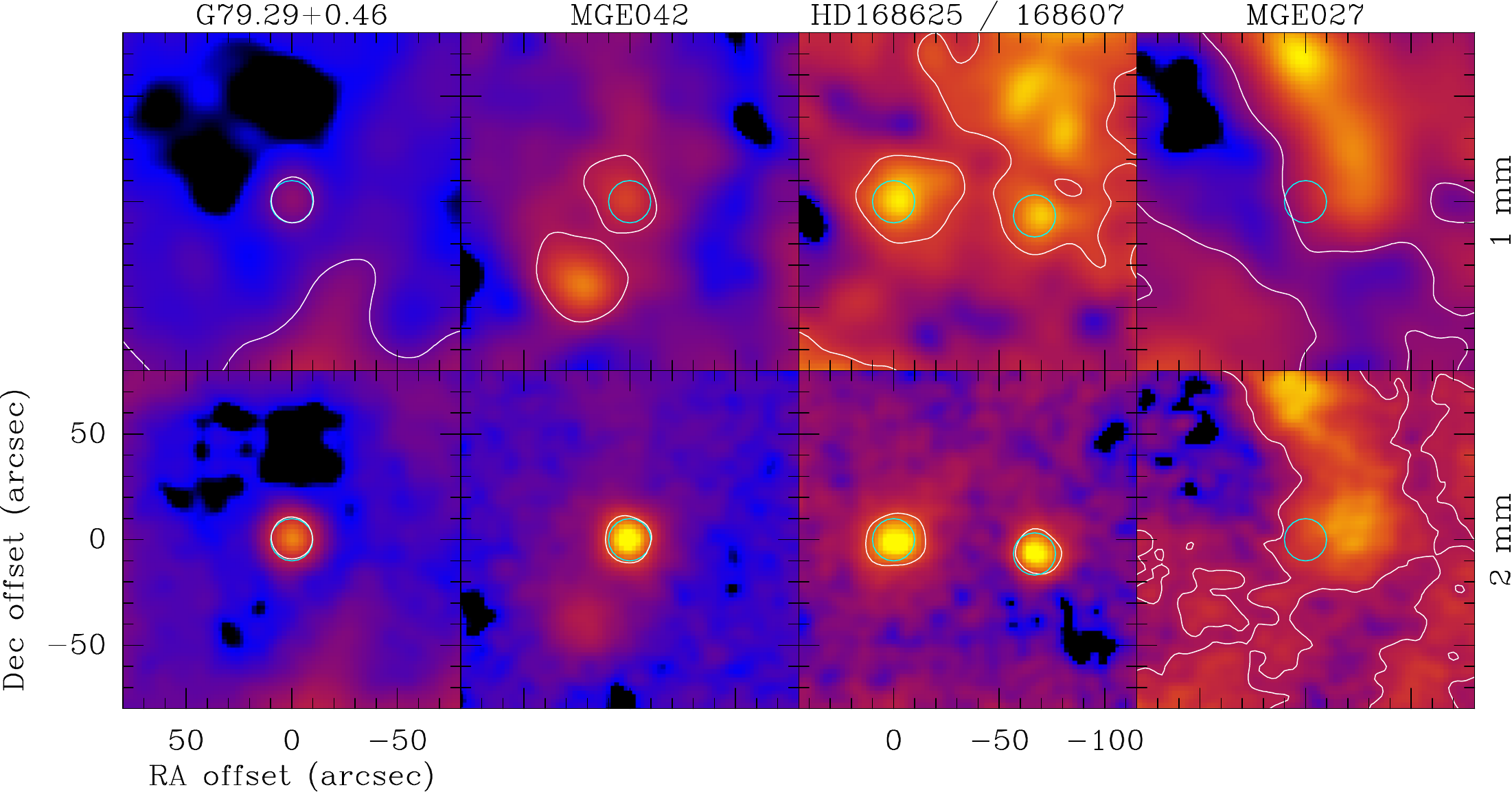} 
\label{cany}
\caption{Details around the LBV stars. Cyan circles indicate the HPBW of the images and are centred at the star positions. White contours correspond to half the flux density at the star position. Note a different range in right ascension for the third field from left, where HD\,168625 is in $(0",0")$ and HD\,168607 is $\approx 70"$ west.}
\end{figure*}

\section{Results}
\label{results}
After combining individual scans and removing the background, we smoothed all images to a common angular resolution of $20"$. This resolution was chosen as a compromise to improve the signal-to-noise ratio without sacrificing too much the level of detail in the maps, and to avoid excessive dilution of the stellar sources. The final images are sketched in Fig.~1. The emission at both bands arises not only from the stars, but also from some circumstellar material and ambient clouds. The white contour is at approximately 5-sigma level, so all the visible features are significant.

The G79.29+0.46 field is dominated by bright and intense emission running from south-east to west. This emission corresponds to an infrared dark cloud (IRDC) linked to DR15, which is known to host active star formation \cite{riv15}, probably induced by the close LBV \cite{riz14,pal14}. The star is detected at both bands, more clearly at 2mm. A shell structure of $\approx 100"$ in radius is also noted, especially at the 2mm band. The NIKA2 shell is clearly correlated with an infrared shell already reported \cite{jim10}. This shell is also bounded by two concentric CO shells related to past mass-loss events \cite{riz08}.

In the MGE042 field we also note a rather extended molecular cloud to the north, not clearly associated with the star. The LBV and a compact cloud $\approx30"$ to the south-east are also clearly identified. It is noteworthy that the star is relatively more intense at 2mm, while the compact cloud looks the opposite. The compact cloud is spatially almost coincident to the densest part of the expanding CO torus \cite{bor19a}.

The LBVs HD\,168625 and HD\,168607 are surrounded by extended features without a clear morphological association. Contrarily to the previous two cases, part of this extended emission is brighter at 2mm. HD168625 seems also immersed within a low-level emission plateau, which is angularly correlated with a bright infrared nebula \cite{oha03}.

The LBV star MGE027 is definitively not detected at the NIKA2 frequencies. Its field is dominated by extended, filamentary, and clumpy material.

The Fig.~2 shows a closer view of the four fields in both bands, in order to disclose the presence of any extended, circumstellar emission around the LBVs. Cyan circles have widths of $20"$ (the angular resolution after smoothing) and are centred on the stars. The contours are half of the peak fluxes on each star. In G79.29+0.46, we see that the contour matches the circle almost perfectly, indicating that this emission arises only from the star and not from possible circumstellar gas or dust. In MGE042, the point source emission arises only at 2mm. The compact cloud at south east is notoriously more intense at 1mm. Both MGE042 and the close compact cloud are immersed within a tenuous plateau, which is more evident at 1mm band. HD\,168625 displays some extended emission in both bands, while HD\,168607 depicts point source emission only at 2mm.

The Table \ref{comps} sketches a quick glance of the features identified. We distinguish five different components or sources of emission: (1) the LBV star; (2) some circumstellar material; (3) more extended emission where the stars are embedded; (4) star-centred shell-like structure; and (5) clouds without a clear morphological relationship to the LBVs. The circumstellar emission appears very close to the stars, while the extended emission (or ``plateau'') is detached from the star emission. There are therefore complex and varied scenarios towards different LBVs, possibly with different composition and excitation mechanisms.

\begin{table*}[hbt]
\centering
\caption{Morphology of the components}
\label{comps}
\begin{tabular}{lcccccccccc}

\hline
\multicolumn{1}{c}{Field} & \multicolumn{2}{c}{Star} & \multicolumn{2}{c}{CSM} & \multicolumn{2}{c}{Plateau} & \multicolumn{2}{c}{Shell} & \multicolumn{2}{c}{Cloud} \\
            & 1mm & 2mm & 1mm & 2mm & 1mm & 2mm & 1mm & 2mm & 1mm & 2mm \\
\hline


G79.29+0.46 & \color{green}{\cmark} & \color{green}{\cmark} & \color{red}{\xmark} & \color{red}{\xmark} & \color{red}{\xmark} & \color{red}{\xmark} & \bf{?} & \color{green}{\cmark} & \color{green}{\cmark} & \color{green}{\cmark}  \\ 
MGE042      & \color{green}{\cmark} & \color{green}{\cmark} & \color{green}{\cmark} & \color{red}{\xmark} & \color{green}{\cmark} & \color{green}{\cmark} & \color{red}{\xmark} & \color{red}{\xmark} & \color{green}{\cmark} & \color{green}{\cmark}  \\ 
HD168625    & \color{green}{\cmark} & \color{green}{\cmark} & \color{green}{\cmark} & \color{green}{\cmark} & \color{green}{\cmark} & \color{green}{\cmark} & \color{red}{\xmark} & \color{red}{\xmark} & \color{green}{\cmark} & \color{green}{\cmark}  \\ 
HD168607    & \color{green}{\cmark} & \color{green}{\cmark} & \bf{?} & \color{red}{\xmark} & \color{green}{\cmark} & \color{green}{\cmark} & \color{red}{\xmark} & \color{red}{\xmark} & \color{green}{\cmark} & \color{green}{\cmark} \\ 
MGE027      & \color{red}{\xmark} & \color{red}{\xmark} & \color{red}{\xmark} & \color{red}{\xmark} & \color{red}{\xmark} & \color{red}{\xmark} & \color{red}{\xmark} & \color{red}{\xmark} & \color{green}{\cmark} & \color{green}{\cmark}  \\ 

\hline
\vspace*{-1mm}

\end{tabular}

{\color{green}{\cmark}}: detected. {\color{red}{\xmark}}: not detected. {\bf ?}: doubtful.

\end{table*}

\section{About the emission mechanisms}
\label{spindex}

NIKA2 frequencies are especially suitable to gather first estimates about the excitation mechanisms of the detected material. Thermal dust is usually relevant at mid-infrared, and steeply decreases when increasing the wavelength. Free-free emission normally from ionized hot gas slightly increases with decreasing frequencies. Depending on the specific physical conditions, both mechanisms may be significant at wavelengths of a few mm.

As we see in the previous section, LBVs are mm-wavelengths continuum sources. Their ejecta are mainly made of ionized atomic gas. When this hot gas moves away from the star it progressively forms molecules, which later may form copious amounts of dust at certain distance. Therefore, we expect important contribution from both free-free gas emission and thermal dust \cite{agl14,agl19}. 

The existence of non-thermal emission, often found in Wolf-Rayet and other massive stars, should not be discarded. The non-thermal emission may have its origin in different processes, such as the shocks of colliding winds in binary systems or the relativistic particle acceleration due to magnetic fields \cite{deb07}. The relative contribution of non-thermal emission at mm-wavelengths is, however, hard to estimate \cite{blo17}.

We can make some guesses about the relative importance of thermal dust, free-free gas and non-thermal processes by computing the spectral index $\alpha$, defined as

\begin{equation}
S_\nu \propto \nu^\alpha
\end{equation}

\noindent where $S_\nu$ is the flux density at the frequency $\nu$. For a spherical expanding {\sc Hii} region we expect $-0.1<\alpha<1.5$ (\cite{tie97} and references therein), while roughly $\alpha=0.6$ is expected for an evolved stellar wind \cite{wri75}, and $\alpha>0.6$ if the wind is radiatively shocked \cite{mon11}. We expect $\alpha=2$ for ideal thermal black body emission, and $\alpha=(2+\beta)$ for a black body modified by the frequency-dependent dust emissivity; $\beta$ is the dust emissivity index, and takes typical values between 1 and 2.7 (\cite{sch10} and references therein).

We computed $\alpha$ in the four fields only for those pixels with flux densities above 2-sigma. The result is shown in Fig.~3. The colour scale runs from $-1$ to 4 in the four panels; blue tones run approximately from $-1$ to 0.6, green tones from 0.6 to 2.3, and dark red tones from 2.3 to 4.

These first results are striking. The spectral indices computed between the two NIKA2 frequencies outstandingly show three separate components, presumably related to different physical conditions or excitation processes. 

The stellar sources are clearly detached from their surroundings, as we can see in the cases of MGE042, HD\,168625 and HD\,168607; those stars and G79.29+0.46 have the lowest $\alpha$, with values well below those corresponding to pure Bremsstrahlung. Such highly negative values of the spectral indices are indicative of a significant contribution of non-thermal processes. 

\begin{figure}[htb]
\centering
\includegraphics[width=1.0\columnwidth]{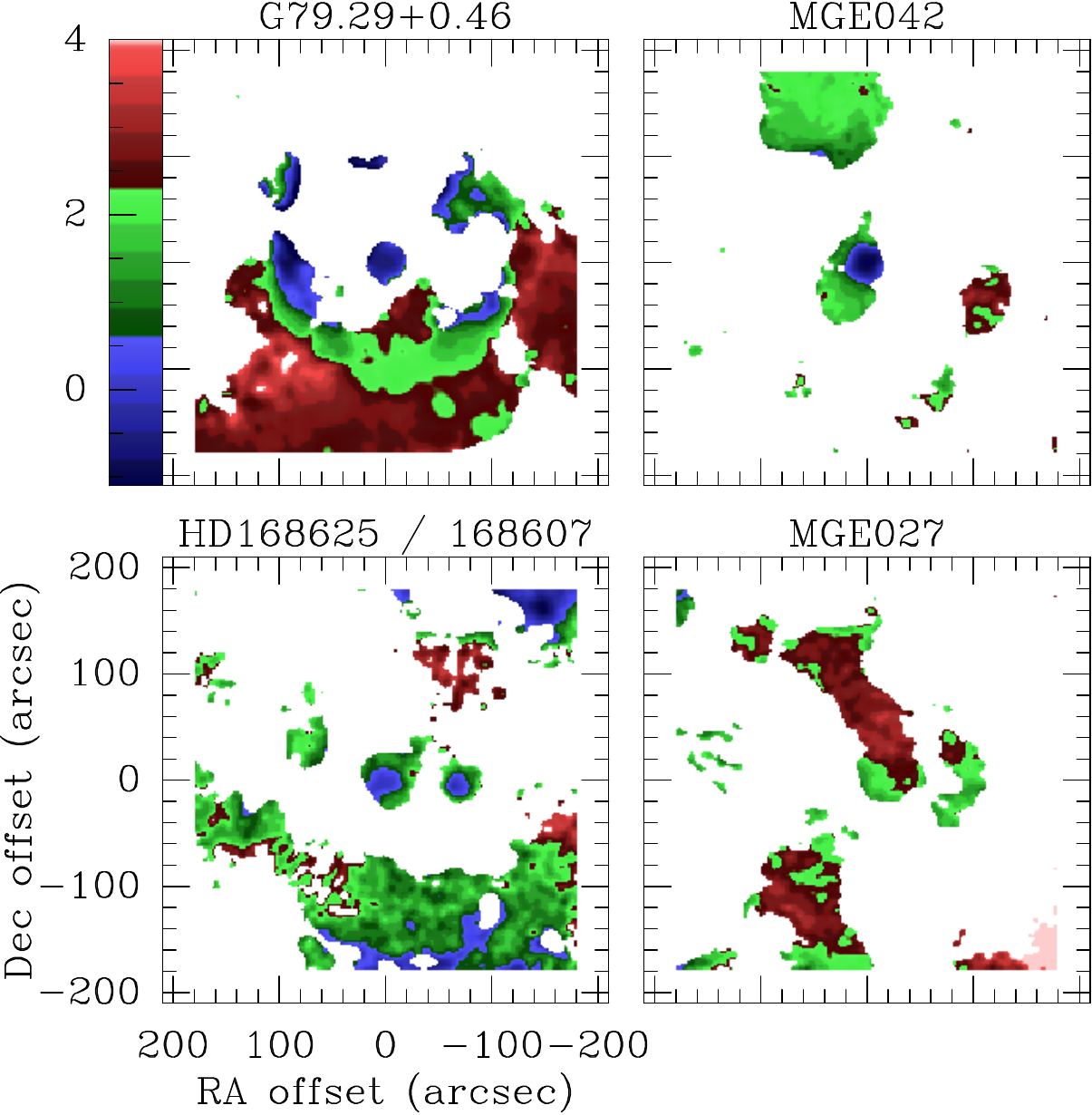} 
\label{alpha}
\caption{Spectral indices of and around the LBV stars observed. The same range is depicted in the four fields.}
\end{figure}

In G79.29+0.46, the bubble enshrouded by the infrared/mm-continuum and CO shells \cite{riz08,jim10} becomes evident, with values of $\alpha$ typical of free-free emission. It is noteworthy the layered structure seen around the $(100",-20")$ position, where it is clearly noted the bubble in blue, the shell affected by a mass-loss process in red and the unaffected IRDC in red. The lowest values of $\alpha$ are consistent with non-thermal emission probably produced by the shocks already discovered \cite{riz14,pal14}.
On the other hand, most of the IRDC (related to DR15) depicts values of $\alpha$ compatible with a cold cloud \cite{sch10}.

The CSM and the plateau identified in MGE042 and HD\,168625/HD\,168607 fields are mostly free-free. In addition, some of these components have values of $\alpha$ compatible with dark clouds.

The MGE027 field is dominated by values of $\alpha$ which are indicative of dark clouds, although with ``skins'' of lower spectral indices (note that this skin is almost absent in the IRDC of the G79.29+0.4 field). This behaviour may indicate that the dark clouds are immersed in a tenuous ambient partly affected by free-free radiation. The non-detection of the star in this case is indeed puzzling. 

\section{Concluding remarks}
\label{concl}
NIKA2 has opened a new road to explore the physical processes associated with the massive star evolution, improving our knowledge of the LBVs themselves and their CSM.

We report the detection of mm-continuum emission in four out of five of the observed LBVs. In addition, some circumstellar material, a shell, and surrounding clouds have also been detected.

First estimates of the spectral indices allowed us to disentangle the nature of the different components. The lowest values of $\alpha$ are found towards the stars and are hardly explained by free-free emission only. We propose that there is some non-thermal contribution to the total flux measured, especially at 2mm. MGE042, HD\,168625, and HD\,168607 are immersed within material having values of $\alpha$ typical of free-free emission. G79.29+0.46 is completely isolated; it is surrounded by a shell of free-free emission, which is clearly interacting with a nearby IRDC. Contrarily, the field of MGE027 is dominated by a cloud apparently driven by thermal dust emission, probably surrounded by tenuous gas.

The results gathered by these observations are relevant, but the whole work is far from being concluded. In some cases (such as G79.29+0.46 at 1mm or HD\,168625 at 2mm) more observations are needed to improve the signal-to-noise ratio. The sample may be enlarged to more LBVs or candidates. Thorough modelling of the thermal and non-thermal processes in each particular case would be made and later analysed. Better angular resolution would be necessary to disclose material close to the star, presumably related to recent ejecta. Finally, the polarimetric capabilities of NIKA2 would be useful to perform polarimetry studies in these scenarios.

\vspace{6mm}
\begin{acknowledgement}
We would like to thank the IRAM staff for their professional support during the campaigns. The NIKA dilution cryostat has been designed and built at the Institut Néel. We acknowledge the crucial contribution of the Cryogenics Group, and in particular Gregory Garde, Henri Rodenas, Jean Paul Leggeri, and Philippe Camus. This work has been partially funded by the Foundation Nanoscience Grenoble and the LabEx FOCUS ANR-11-LABX-0013. This work is supported by the French National Research Agency under the contracts "MKIDS" and "NIKA" and in the framework of the "Investissements d’avenir” program (ANR-15-IDEX-02). This work has benefited from the support of the European Research Council Advanced Grant ORISTARS under the European Union’s Seventh Framework Programme (Grant Agreement no. 291294). J.R.R. acknowledges support from Spanish grants ESP2015-65597-C4-1-R (MINECO/FEDER, UE) and ESP2017-86582-C4-1-R (AEI/FEDER, UE). A.R. acknowledges support from Spanish grant AYA2015-66211-C2-2 (MINECO/FEDER, UE).
\end{acknowledgement}

\end{document}